\documentclass[a4paper,10pt,prd,aps,showpacs,superscriptaddress]{revtex4}
\usepackage{amsmath, amsfonts, amssymb, float, appendix, multirow, longtable}
\usepackage{graphicx}
\usepackage{color}
\usepackage{hyperref}
\usepackage{mathrsfs}
\usepackage{subfigure}

\newcommand{\sfrac}[2]{{\textstyle{#1\over#2}}}
\newcommand{\dd}{\text{D}}

\begin{document}

\title{Cosmic Electromagnetic Fields due to Perturbations in the Gravitational Field}


\author{Bishop Mongwane}
\email{astrobish@gmail.com} 
\affiliation{Astrophysics Cosmology \& Gravity Center, and Department of Mathematics \& Applied Mathematics,
  University of Cape Town, 7701 Rondebosch, South Africa}

\author{Peter K. S. Dunsby}
\email{peter.dunsby@uct.ac.za}
\affiliation{Astrophysics Cosmology \& Gravity Center, and Department of Mathematics \& Applied Mathematics,
  University of Cape Town, 7701 Rondebosch, South Africa}
\affiliation{South African Astronomical Observatory, Observatory
7925, Cape Town, South Africa}

\author{Bob Osano}
\email{bob.osano@uct.ac.za}
\affiliation{Astrophysics Cosmology \& Gravity Center, and Department of Mathematics \& Applied Mathematics,
  University of Cape Town, 7701 Rondebosch, South Africa}

\date{\today}

\begin{abstract}
We use non-linear gauge-invariant perturbation theory to study the interaction of an inflation produced seed magnetic
field with density and gravitational wave perturbations in an almost Friedmann-Lema\^itre-Robertson-Walker (FLRW)
spacetime with zero spatial curvature. We compare the effects of this coupling under the assumptions of poor conductivity, perfect conductivity and the case
where the electric field is sourced via the coupling of velocity perturbations to the seed field in the ideal
magnetohydrodynamic (MHD) regime, thus generalizing, improving on and correcting previous results.
We solve our equations for long wavelength limits and numerically integrate the resulting equations to generate power spectra for the electromagnetic field variables,  showing where the modes cross the horizon. We find that the interaction can seed Electric fields with non-zero curl and that the curl of the electric field 
dominates the power spectrum on small scales, in agreement with previous arguments.
\end{abstract}
\maketitle
\section{Introduction}
Large scale magnetic fields of varying amplitudes are present in entire galaxy clusters,
individual galaxies and high redshift condensations. Such fields are observed on
characteristic scales of $\sim 1$ Mpc and are of micro-Gauss strength, $10^{-7}-10^{-5}$ G
\cite{Kronberg:1993vk,Widrow:2002ud}.
Despite their ubiquity, their origin is still a mystery. There are literally tens of candidate mechanisms proposed to
explain the origin and evolution of such fields, spanning different
theories of physics \cite{Battaner:2000kf}. It is now widely believed that the structure of magnetic fields in spiral
galaxies is consistent with the dynamo amplification mechanism. The dynamo mechanism can produce amplification factors 
of up to $\sim 10^{8}$ but requires a seed field in order to operate and thus cannot explain the origin of magnetic
fields. Additionally, adiabatic contraction of magnetic flux lines during structure formation can enhance 
galactic fields by a factor of $\sim 10^{3}$.

Among the physical mechanisms proposed to explain the origin of the seed field is one due to Harrison
\cite{harrison:1970}.
This mechanism rests on the fact that non-zero vorticity in the pre-recombination photon-baryon plasma can generate weak
magnetic fields of about $\sim 10^{-25}$ G. However, vorticity
is not a generated mode at first order in perturbation theory and has to be put in as an initial condition. Second order
treatments of the pre-recombination plasma in terms of a Kinetic theory description has also been used to generate the
required seed fields \cite{Gopal:2004ut,Fenu:2010kh,Kobayashi:2007wd,Ichiki10022006,Takahashi:2005nd,Maeda:2008dv}. The key idea is a preferential
Thompson scattering of photons off free electrons, over the scattering off protons (the scattering off protons is
suppressed by a factor
$(m_{e}/m_{p})^2$) which induces differences in the proton and
electron velocity fields. Electric fields are then
induced to counter charge separation between the electrons and protons.
The generated electric fields will then feed in the magnetic induction equation to generate magnetic fields at second
order in perturbation theory. The
photon anisotropic stress also couples to the electron velocities and contributes to the magnetic field sources.
In addition, other arguments relying on electroweak phase transitions
\cite{Hogan:1983zz,Caprini:2009yp}, topological defects \cite{Hollenstein:2007kg}, velocity perturbations \cite{Betschart:2003bn}
etc. have been proposed as candidate mechanisms. 
The generated fields, however, are usually too weak to leave
any detectable
imprint on the CMB \cite{Fenu:2010kh}. This is not surprising given the form of the fluid quantities of a magnetic
field. In particular,
the
energy density $\mu_{B}=B^{2}/2$, the isotropic pressure $p_{B} = B^{2}/6$ and the anisotropic pressure
$\Pi_{ab}=B_{\langle a}B_{b \rangle}$ of a field generated at second order will manifest at fourth order in perturbation
theory, which is not relevant for CMB anisotropies.

In addition to meeting the right strengths, the generated fields must be of the right scale to match those observed
today. One of the problems of 
primordial generation mechanisms in general is that although some may reach the required strengths, they are causal in
nature. This
 means that their coherence scales cannot exceed the Hubble scale during the time of magnetic field generation. 
By comparison, the galactic scale today is well outside the Hubble scale at such early epochs. Moreover, the small
scale fields i.e, those that are already sub-horizon before matter-radiation equality cannot reach the
recombination epoch due to micro-physical mechanisms such as magnetic and photon diffusion processes
\cite{Battaner:2000kf}. 

Inflation and other pre-Big Bang models capable of causally producing super horizon perturbations are often invoked to
circumvent this scale problem. However, the residual magnetic fields surviving the exponential expansion accompanying
many inflationary models are thought to be too weak to be of cosmological relevance\footnote{This is not a
generic feature of all Friedmann universes however. It is possible to `preserve' primordial magnetic fields in an
open Friedmann universe; the hyperbolic geometry can slow down the adiabatic decay of the field leading to
superadiabatic amplification \cite{Barrow:2008jp,Tsagas:2005nn}, see also \cite{Adamek:2011hi}.}. New physics often has
to be introduced such as
exotic couplings of the electromagnetic field to other fields such as the dilaton field to avoid
the accompanying 
exponential dilution of the magnetic fields \cite{gasperini:1995}. The primordial fields are
also constrained by the fact that the anisotropic stress of the produced magnetic fields contains a spin-2 component
and will result in an
overproduction of gravitational waves at horizon crossing which is inconsistent with standard Big Bang Nucleosynthesis
constraints \cite{Caprini:2001nb,Caprini:2005ed}.

Apart from studying the generation of magnetic fields, one can also study interactions of a pre-existing
magnetic field with gravitational degrees of freedom. This is often studied in the context of amplification of the seed
magnetic field or gravitational wave detection. Much progress has been made in this area
\cite{Marklund:2000zs,Tsagas:2001ak,Betschart:2005iu,Fenu:2008vs}. Most of these studies however have been restricted to 
focusing on the interaction of magnetic fields with tensor perturbations; In this work we revisit and extend the work
presented in \cite{Marklund:2000zs,Betschart:2005iu,Zunckel:2006cs}, to include scalar perturbations in the matter fluctuations.

When using perturbation theory about a FLRW background to study the interaction, one is immediately faced with
the problem of how to embed the seed magnetic field into the background. The isotropy of the FLRW spacetime does not readily allow
for any direction preference that may be introduced by a vector field. There are several ways to handle this and we
mention briefly just three of them. One can
treat the seed magnetic field as a zeroth order quantity, subject to the assumption that the energy density of the field
be small compared to the energy density of matter $B^{2} \ll \mu $ and that the anisotropic stress is negligible
$\Pi_{ab}=B_{\langle a}B_{b \rangle}\approx 0$. With
these approximations, the energy density of the magnetic field cannot alter the gravitational dynamics of the
background spacetime; this
approach is often referred to as the weak-field approximation. Another approach is to treat the seed field as a
statistically homogeneous and isotropic random field with $\langle \mathbf{B} \rangle =0$ but $\langle B^{2} \rangle
\neq 0$ and so, the seed field does not introduce any directional dependence in the background
spacetime. One can then easily employ statistical methods to quantify the field's behavior.
Another
possibility is to leave the background spacetime untouched but treat the seed field as a first order perturbation, using
a two
parameter approximation scheme to characterize the perturbations in the electromagnetic and gravitational field; this
is the approach we adopt in this work.

One can go a long way in comparing the different perturbation schemes. For example, in the weak-field approximation, the
induced magnetic field will be at first order, a well understood regime in perturbation theory. While in the two
parameter case, the induced field will be at second order \footnote{For the purposes of this argument, we refer to
perturbations of order $\epsilon_{g} \epsilon_{\tilde{B}}$ as second order, with evident misuse of terminology. In the
rest of the work, we will refer to quantities of this order simply as non-linear and reserve the designation `second
order' to quantities of order $\epsilon_{g}^{2}$ and $\epsilon_{B}^{2}$}, a regime that is not so well developed.
Nevertheless, for the purposes of our work, the two approaches are mathematically equivalent. The apparent differences
between them is as a result of 
relabeling of spacetimes, i.e. `First order' in the weak-field approximation corresponds to `second order' in the two
parameter case. Indeed, Maxwell's equations and thus the Einstein-Maxwell system takes the same mathematical form in
both of these approaches. They both use the machinery of relativistic perturbation theory and are thus prone to gauge
issues, see \cite{Hortua:2011mn,Tsagas:1999tu} for example.

The present article is structured as follows: we present details of our perturbative framework in
\S~\ref{sec:framework}. After a presentation
of the interaction equations in \S\ref{sec:interaction}, we  present the derivation of the equations describing
the induction of EM fields in \S\ref{sec:efield} and \ref{sec:bfield} for a general current and a note on how to
evaluate the induced electrical current in \S\ref{sec:current}. We present the power spectra of the induced magnetic
field
variable in \S\ref{sec:powerspec} and finally a summary in \ref{sec:conclusion}.
We employ the 1+3 covariant approach to perturbation theory \cite{ellis:1971} and follow \cite{gravitation:1973} by
adopting the more geometrically motivated metric
signature $(-+++)$ and we use 
geometrized units $8\pi G=c=1$, where $G$ is the gravitational constant and $c$ is the speed of light in vacuum.

\section{Preliminaries}
\subsection{1+3 spacetime splitting}

One of the nice aspects of the 1+3 covariant approach to General Relativity (GR) is that the underlying dynamical equations               
have a stronger appeal from a physical point of view, as compared to the quasi-linear, second-order partial
 differential equation form, which the EFE take in the metric based approach.
 
The approach is based on a 1+3 decomposition of geometric quantities with respect to a fundamental four velocity
$u^{a}$. 
\begin{equation}
u_{a}=\frac{dx^{a}}{d\tau}, \qquad u_{a}u^{a}=-1\;,
\end{equation}
where $x^{a}$ are general coordinates and $\tau$ measures the proper time along the world line.
The key equations governing the full structure of the spacetime are derived from the Ricci and the once and twice
contracted Bianchi identities applied to the 4-velocity vector \cite{Ellis:1998ct}.
This splitting uniquely defines two projection tensors
\begin{eqnarray}
U^{a}_{\phantom{a}b} = -u^{a}u_{b} \qquad &\Rightarrow& \qquad  U^{a}_{\phantom{a}c}
U^{c}_{\phantom{a}b}=U^{a}_{\phantom{a}b},\; U^{a}_{\phantom{a}a}=1,\; U_{ab}u^{b}=u_{a} \;,\\
h_{ab} = g_{ab}+u_{a}u_{b}\qquad &\Rightarrow& \qquad h^{a}_{\phantom{a}c} h^{c}_{\phantom{a}b} =
h^{a}_{\phantom{a}b},\; h^{a}_{\phantom{a}a}=3, \; h_{ab}u^{b}=0\;,
\end{eqnarray}
which project along and orthogonal to the 4-velocity $u^{a}$. We define two projected covariant derivatives, the
convective time derivative along $u^{a}$ and the spatially projected covariant derivative 
\begin{equation}
\dot{Q}^{a\cdots b}_{\phantom{a\cdots b} c\cdots d} \equiv u^{e}\nabla_{e} Q^{a\cdots b}_{\phantom{a\cdots
b} c\cdots d}  \qquad\text{and}\qquad
 \dd_{e} Q^{a\cdots b}_{\phantom{a\cdots b} c\cdots d}\equiv
h^{a}_{\phantom{a}p} \cdots h^{b}_{\phantom{q}q}\, h^{r}_{\phantom{a}c}\cdots h^{s}_{\phantom{a}d}\,
h^{f}_{\phantom{e}e}\nabla_{f} Q^{p\cdots q}_{\phantom{a\cdots b} r\cdots s}
\end{equation}
respectively.
The basic equations are then characterized by the irreducible parts of the first covariant derivative of $u_{a}$
\begin{equation}
\nabla_{a}u_{b}=-u_{a}\mathcal{A}_{b} + \dd_{a}u_{b}=-u_{a}\mathcal{A}_{b}+\frac{1}{3}\Theta
h_{ab}+\sigma_{ab}+\omega_{ab}\;,
\end{equation}
where $\mathcal{A}_{b}=u^{a}\nabla_{a}u_{b}$ is the relativistic acceleration vector representing the effect of
inertial forces on the fluid;  $\dd_{a}u^{a}=\Theta$ is the rate of volume expansion; $\sigma_{ab}=\dd_{\langle a}u_{b
\rangle}$ is the
symmetric trace-free rate of shear tensor, describing the rate of distortion of the fluid flow;
$\omega_{ab}=\dd_{[a}u_{b]}$ is the antisymmetric vorticity tensor, describing the rigid rotation of the fluid relative
to a non-rotating frame.

\subsection{FLRW background}

We choose as our background the FLRW models, which are spatially homogeneous and isotropic. Thus, relative to
the congruence $u_{a}$,  the kinematical variables have to be locally isotropic, which implies the vanishing of the
4-acceleration $\dot{u}_{a}=0$, the rate of shear $\sigma_{ab}=0$ and the vorticity vector $\omega_{a}=0$. Spatial
homogeneity
implies that the spatial gradients of the energy density
$\mu$, pressure $p$, and the expansion $\Theta$ vanish, i.e $\dd_{a}\mu=\dd_{a}p=\dd_{a}\Theta=0$. Moreover, the FLRW
spacetime is characterized by a perfect fluid matter tensor, i.e $\pi=q_{a}=0$. These restrictions imply that the
spacetime is conformally flat, i.e the electric and magnetic parts of the Weyl tensor vanish, $E_{ab}=H_{ab}=0$.
This leads to the key background equations, the energy conservation equation 
\begin{equation}
\label{eq:energy}
\dot{\mu}=-(1+w)\Theta \mu,
\end{equation}
the Raychaudhuri equation
\begin{equation}
\label{eq:raychaudhuri}
\dot{\Theta}=-\frac{1}{3}\Theta^{2}-\frac{1}{2}\mu(1+3w)+\Lambda,
\end{equation}
where $w=p/\mu$ and the Friedmann equation
\begin{equation}
\mu+\Lambda=\frac{1}{3} \Theta^{2}+\frac{3K}{a^{2}}.
\end{equation}


\section{Perturbative Framework}
\label{sec:framework}

As already mentioned, a FLRW spacetime cannot readily host magnetic fields, as their anisotropic stresses
${\Pi_{ab}=\tilde{B}_{\langle a}\tilde{B}_{b \rangle}\neq 0}$ will break the isotropy. We thus treat the background
magnetic field $\tilde{B}_{a}$ as a first order perturbation to the isotropic spacetime. This lends the energy
density, the isotropic and anisotropic pressure of the field to second order in perturbation theory.

We then proceed by adopting a two parameter perturbative
framework \cite{Bruni:2002sma,Sopuerta:2003rg, Clarkson:2003mp,Passamonti:2004kz,Passamonti:2004je}. Fundamentally, this
consists of
separately parametrizing the gravitational and Maxwell field perturbations in two expansion parameters $\epsilon_{g}$
and $\epsilon_{\tilde{B}}$, representing the
amplitudes of the gravitational and electromagnetic field perturbations, respectively 
\cite{Clarkson:2003mp,Betschart:2005iu,Zunckel:2006cs}. Using this parametrization, any quantity
$Q^{...}_{\phantom{...}...}$ in the physical spacetime can be expanded in the form,
\begin{equation}
Q^{...}_{\phantom{...}...} = \epsilon_{g}^{0} \epsilon_{\tilde{B}}^{0}
\stackrel{\scriptscriptstyle{(0,0)}}{Q^{...}_{\phantom{...}...}} + \epsilon_{g}^{1} \epsilon_{\tilde{B}}^{0}
\stackrel{\scriptscriptstyle{(1,0)}}{Q^{...}_{\phantom{...}...}} + \epsilon_{g}^{0} \epsilon_{\tilde{B}}^{1}
\stackrel{\scriptscriptstyle{(0,1)}}{Q^{...}_{\phantom{...}...}} + \epsilon_{g}^{1} \epsilon_{\tilde{B}}^{1}
\stackrel{\scriptscriptstyle{(1,1)}}{Q^{...}_{\phantom{...}...}} + \mathcal{O}(\epsilon_{g}^{2}\;,
\epsilon_{\tilde{B}}^{2}) \;,
\end{equation}
where the first term on the right represents the background term; the first and second terms represent the
first order gravitational and electromagnetic perturbations respectively; the fourth term represents the
non-linear coupling we're looking to investigate; the higher order terms represent self-coupling terms of order
$\epsilon_{g}^{m}$ and $\epsilon_{\tilde{B}}^{n}$, $m,n \geq 2$. In general, terms describing the coupling
will be of the form 
$\epsilon_{g}^{m} \epsilon_{\tilde{B}}^{n}$, where, in this work, we restrict the perturbative order to 
$\mathcal{O} (\epsilon_{g}^{1}\epsilon_{B}^{1})$ and therefore neglect terms of
order $\mathcal{O}(\epsilon_{g}^{2}\epsilon_{B}^{1})$, $\mathcal{O} (\epsilon_{g}^{1}\epsilon_{B}^{2})$ and higher,
resulting from the self-coupling of the fields; this includes gravitational couplings with the magnetic anisotropy 
$\Pi_{ab}= -\tilde{B}_{\langle a} \tilde{B}_{b\rangle}$, leading to $\mathcal{O} (\epsilon_{g}^{1}\epsilon_{B}^{2})$
terms. We will generally
refer to quantities of order
$\mathcal{O}(\epsilon_{g}^{1}\epsilon_{B}^{1})$ simply as non-linear and reserve the designation `second order' for
terms that
 are of order $\epsilon_{g}^{2}$ and $\epsilon_{\tilde{B}}^{2}$. As in \cite{Clarkson:2003mp,Betschart:2005iu,
Zunckel:2006cs}, one
can visualize this framework as a hierarchy of spacetimes to label the different
perturbative orders. 

We make the common assumption in the literature that the perturbed spacetimes have the same manifold as the background
spacetime i.e. we consider the perturbations as fields propagating on the background spacetime 
\cite{stewart:1990,Sopuerta:2003rg}. In
this treatment, therefore, we restrict the possibility that the perturbations may alter the differential structure of
the background manifold and so we neglect issues of backreaction.

We're also interested in studying this coupling in a gauge-invariant manner. The gauge problem in relativistic
perturbation theory has been dealt with in the literature, see for example 
\cite{stewart:1974,ellis:1989,Sonego:1997np,Sopuerta:2003rg,Bruni:1996im}. The Stewart \& Walker Lemma
\cite{stewart:1974} serves as a basis for the generalization of gauge invariance to
arbitrary order \cite{Sonego:1997np,Sopuerta:2003rg}. It follows that a quantity $Q$ is gauge invariant at order
$\mathcal{O}(\epsilon_{g}^{m} \epsilon_{\tilde{B}}^{n})$ if and 
only if $Q^{\scriptscriptstyle{(0)}}$ and its perturbations of order lower than $\mathcal{O}(\epsilon_{g}^{m}
\epsilon_{\tilde{B}}^{n})$ are
either vanishing, or a constant scalar or a combination of Kronecker deltas with constant coefficients
\cite{Bruni:2002sma,Sopuerta:2003rg}.

Since the interaction terms are of order $\mathcal{O}(\epsilon_{g}^{1}\epsilon_{B}^{1})$ we have that the
induced magnetic field $B_{a}$ will be of the same order; we also assume that the electric field $E_{a}$ will be of
the same order as the induced magnetic field. Clearly $B_{a}$ does not satisfy the criteria for gauge invariance at
$\mathcal{O}(\epsilon_{g}^{1}\epsilon_{\tilde{B}}^{1})$ since
it is neither vanishing nor a constant scalar at $\mathcal{O}(\epsilon_{g}^{0} \epsilon_{\tilde{B}}^{1})$. To this end,
we make use of the same auxiliary variable $\beta_{a}=\dot{\tilde{B}}_{a}+\frac{2}{3}\Theta \tilde{B}_{a}$ identified in \cite{Clarkson:2003mp,Betschart:2005iu,
Zunckel:2006cs}. We do
not however integrate $\beta_{a}$ to recover the gauge-dependent magnetic field, but treat it as the fundamental
variable whose deviation from zero quantifies deviation from the adiabatic decay of the magnetic field.

\section{The Einstein-Maxwell system}

The Einstein-Maxwell equations~\ref{eq:max} contain terms that couple the electromagnetic fields to the gravitational
fields. These can be written at $\mathcal{O}(\epsilon_{g}^{1}
\epsilon_{\tilde{B}}^{1})$ by discarding higher order terms. This results in the two propagation equations,
\begin{eqnarray}
\label{eq:induction_B}
\dot{B}_{\langle a \rangle} +  \frac{2}{3}\Theta B_{a} &=&  \sigma_{ab} \tilde{B}^{b} - \mathrm{curl}\; E_{a}\;,\\
\label{eq:induction_E}
\dot{E}_{\langle a \rangle} +\frac{2}{3}\Theta E_{a} &=&  \mathrm{curl}\; B_{a} + \epsilon_{abc} \mathcal{A}^{b}
\tilde{B}^{c} - \mathcal{J}_{a} \;,
\end{eqnarray}
subject to the constraints, $\dd_{a} E^{a} =0= \dd_{a} B^{a}$. Following \cite{Clarkson:2003af} we make the following
comments:
(i) The magnetic field $\tilde{B}_{a}$ appearing in the equations~\ref{eq:induction_B} and~\ref{eq:induction_E}
multiplied by the gravitational variables should not be the same as the $B_{a}$ appearing alone. The variable
$B^{a}$ is a mixture of linear and non-linear quantities (the seed magnetic field and
the induced field) while the terms involving $\tilde{B}^{a}$ are a product of first order quantities. One has to keep
this in mind when integrating the equations.
(ii) The system is not gauge-invariant as already mentioned in \S~\ref{sec:framework}. This can be attributed to the
mixture of linear and non-linear terms in the system. In the covariant approach to perturbation theory, the solution of
perturbed differential operators is never sought, one can get around this by making sure that the differential
operators involved operate on quantities of the corresponding perturbative order.

In an attempt to cast it in a consistent and gauge invariant manner, we introduce the following non-linear variables:
The fundamental variable $\beta_{a}$ measuring deviation from adiabatic decay, $I_{a}$ describing the interaction with
shear distortions and $\xi_{a}$ describing interaction with density perturbations. These are defined as,
\begin{equation}
\label{eq:defs}
\beta_{a}=\dot{B}_{\langle a \rangle} + \frac{2}{3}\Theta B_{a} \;, \qquad I_{a}=\sigma_{ab}\tilde{B}^{b} \qquad
\text{and} \qquad \xi_{a} =\epsilon_{abc} \mathcal{A}^{b}
\tilde{B}^{c}
\end{equation}
and results in the following system,
\begin{eqnarray}
\label{eq:induction_B}
\beta_{a} &=&  I_{a} - \mathscr{E}_{a}\;, \\
\label{eq:induction_E}
\dot{E}_{\langle a \rangle} +\frac{2}{3}\Theta E_{a} &=&  \mathscr{B}_{a} + \xi_{a} - \mathcal{J}_{a}\;,
\end{eqnarray}
where we have written $\mathrm{curl}\; E_{a}=\mathscr{E}_{a}$ and $\mathrm{curl}\; B_{a}=\mathscr{B}_{a}$ for brevity.

\section{The linear equations}
\subsection{The linear magnetic field: $\mathcal{O}(\epsilon_{\tilde{B}})$}
\label{sec:bfield}

We treat the seed
magnetic field
as a first order perturbation to the spacetime. The seed field may have its origins in inflation or other mechanisms 
based on string cosmology, in which electromagnetic vacuum fluctuations are amplified due to a dynamical dilaton or an
inflaton field \cite{gasperini:1995}.
We assume that at order $\mathcal{O}(\epsilon_{g}^{0} \epsilon_{\tilde{B}}^{1})$ the electric fields are small
compared to the magnetic fields, i.e $E^{2}\ll B^{2}$. Thus, in the absence of diffusive losses or amplification, the
induction equation \ref{eq:induction_B} takes the frozen-in form,
\begin{equation}
\label{eq:induction}
\dot{\tilde{B}}_{\langle a \rangle}+\frac{2}{3}\Theta \tilde{B}_{a}=0\;,
\end{equation}
regardless of the equation of state or plasma properties of the cosmic fluid. It follows then that the magnetic field
decays adiabatically as $\tilde{B}_{a}\propto a^{-2}$, where $a$ is
the cosmological scale factor. This adiabatic decay arises from the expansion of the Universe which conformally dilutes
the field lines due to flux conservation. 
The frozen-in condition ~\ref{eq:induction} does not discriminate between homogeneous ($\dd_{a}\tilde{B}_{b}=0$) and
inhomogeneous ($\dd_{a}\tilde{B}_{b}\neq 0$) magnetic fields. For an
inhomogeneous field the spatial gradients of the seed magnetic field $\dd_{b} \tilde{B}_{a}$ are of the same order
as $\tilde{B}_{a}$ and
evolve as $\dd_{b}B_{a}\propto a^{-3}$. 

\subsection{Gravitational perturbations: $\mathcal{O}(\epsilon_g)$}

The Weyl tensor $C_{abcd}$ represents the free gravitational field, enabling gravitational action at a distance.
In analogy with splitting the Maxwell field tensor $F_{ab}$ into a magnetic and an electric field, $C_{abcd}$ can
be split covariantly into a `magnetic' part $H_{ab}=\frac{1}{2}\epsilon_{ade}C^{de}_{\phantom{bc}bc} u^{c}$ and an
`electric' part $E_{ab} = C_{abcd}u^{c}u^{d}$. The electric part of the Weyl tensor describes tidal effects, akin to
the tidal tensor associated with the Newtonian potential, while the magnetic part describes the propagation of
gravitational radiation. The Weyl tensor vanishes in the conformally flat FLRW spacetime and so $E_{ab}$ and $H_{ab}$
 are covariant first order gauge invariant (FOGI) quantities in the Weyl curvature.
We also define the FOGI variables $\mathcal{X}_{a}=a\dd_{a}\mu$ and $\mathcal{Z}=a\dd_{a}\Theta$ to characterize density
perturbations.
Now, the system governing gravitational perturbations is given by the following propagation
equations \footnote{Equation~\ref{eq:propshear}, is
obtained from the Ricci identities applied to the whole spacetime;~\ref{eq:propMagH} and \ref{eq:propelecW} are
obtained from the once
contracted Bianchi identities;~\ref{eq:density} and \ref{eq:expansion} are obtained by taking 
spatial gradients of the energy conservation and Raychaudhuri equation, respectively},

\begin{eqnarray}
\label{eq:propshear}
\dot{\sigma}_{\langle ab \rangle}+\frac{2}{3}\Theta \sigma_{ab} &=& \dd_{\langle a} \mathcal{A}_{b \rangle} -E_{ab}\;, \\
\label{eq:propMagH}
\dot{H}_{\langle ab \rangle}+\Theta H_{ab}&=& -\mathrm{curl}\;E_{ab}\;, \\
\label{eq:propelecW}
\dot{E}_{\langle ab \rangle}+\Theta E_{ab} &=& \mathrm{curl}\;H_{ab}-\frac{1}{2}\mu (1+w)
\sigma_{ab}\;, \\
\label{eq:density}
\dot{\mathcal{X}}_{\langle a \rangle} - \Theta w \mathcal{X}_{a} &=& - (1+w)\mathcal{Z}_{a}\;, \\
\label{eq:expansion}
\dot{\mathcal{Z}}_{\langle a \rangle} + \frac{2}{3}\Theta \mathcal{Z}_{a} &=& -\frac{1}{2}\mu \mathcal{X}_{a} -
\frac{w}{3(1+w)}\left(-\frac{1}{3}\Theta^{2}+\mu + \Lambda \right)\mathcal{X}_{a} -\frac{w}{1+w}
\dd^{2}\mathcal{X}_{a}\;.
\end{eqnarray}

In addition to the
propagation equations above, the following constraints have to be satisfied,
\begin{equation}
\label{eq:constraints}
a\dd^{c}\sigma_{bc}=\frac{2}{3} \mathcal{Z}_{b}, \qquad aD^{c}E_{bc}=\frac{1}{3}\mu \mathcal{X}_{b}
\qquad\text{and}\qquad
H_{ab}=\text{curl}\;\sigma_{ab}\;,
\end{equation}
where we have set the vorticity to zero ($\omega_{a}=0$), see also \cite{Clarkson:2003af}. Note that at first order in
gravitational perturbations, the only source of vector modes 
is the vorticity $\omega_{a}$; since, we neglect the effects of
vorticity, $\omega_{a}=0$, all the vector modes vanish.
The shear tensor $\sigma_{ab}$ can then be irreducibly split into scalar and tensor contributions as \cite{Clarkson:2011td}
\begin{equation}
\sigma_{ab}=\sigma^{\scriptscriptstyle{S}}_{ab} + 
\sigma^{\scriptscriptstyle{T}}_{ab} \quad \mathrm{where}\quad
\mathrm{curl}\;\sigma^{\scriptscriptstyle{S}}_{ab}=0,\quad
\mathrm{and} \quad
\dd^{a}\sigma^{\scriptscriptstyle{T}}_{ab}=0\;.
\end{equation}
 The pure tensor modes can be used to characterize
gravitational waves \cite{Dunsby:1998hd}. The scalar part of the shear couples to density perturbations and is
related to the clumping of matter via the constraints ~\ref{eq:constraints}.

By differentiating~\ref{eq:propshear} and using \ref{eq:propelecW} and one of the constraints \ref{eq:constraints} to
substitute for $E_{ab}$ and $H_{ab}$, one arrives at a forced wave equation for the shear,
\begin{eqnarray}
\label{eq:shearscal}
\ddot{\sigma}_{\langle ab \rangle} -\dd^{2} \sigma_{ab} +\frac{5}{3} \Theta \dot{\sigma}_{\langle ab \rangle}+
\left[\frac{1}{9} \Theta^{2}+\frac{1}{6} \mu-\frac{3}{2} p+\frac{5}{3}\Lambda\right]\sigma_{ab}=
-\frac{w}{a^{2}(1+w)}\left[\dot{\mathcal{X}}_{ab} +\frac{1}{3}\Theta \mathcal{X}_{ab} \right]\;,
\end{eqnarray}
where $\mathcal{X}_{ab}=-(1+w)a^{2}\dd_{\langle a}\mathcal{A}_{b \rangle}/w = a \dd_{\langle a}\mathcal{X}_{b \rangle}$.
We need an evolution equation for $\mathcal{X}_{ab}$ in order to close equation \ref{eq:shearscal}.
One can start from ~\ref{eq:density} and ~\ref{eq:expansion} to write a wave equation for $\mathcal{X}_{a}$ then taking
the comoving spatial gradient of the resulting wave equation will yield the following,
\begin{equation}
\ddot{\mathcal{X}}_{ab}-w\dd^{2}\mathcal{X}_{ab}-\left(w-\frac{2}{3}\right)\Theta
\dot{\mathcal{X}}_{ab}+\frac{1}{2}\mu(3w+1)(w-1)\mathcal{X}_{ab}-2w\Lambda \mathcal{X}_{ab}=0\;.
\end{equation}

In including scalar perturbations, we have
explicitly coupled the shear tensor to density perturbations. This shows that density gradients source distortions
in the Weyl curvature and vice versa. Hence, knowing the shear allows one to compute density gradients and knowing
density
gradients one can compute the scalar part of the shear \cite{bruni:1992}.

\section{The interaction: $\mathcal{O}(\epsilon_{g} \epsilon_{\tilde{B}})$}
\label{sec:interaction}

The Maxwell fields couple to Weyl curvature through the shear term and 
density perturbations through the acceleration terms and the non-linear
identity~\ref{eq:identity}.
In the case of pure tensor modes in the shear tensor, the interaction variable
$I_{a}=\sigma^{\scriptscriptstyle{T}}_{ab}\tilde{B}^{b}$ was
shown to satisfy a closed wave equation, for
both a homogeneous \cite{Betschart:2005iu} and an inhomogeneous \cite{Zunckel:2006cs} seed field $\tilde{B}_{a}$.
Here, we include contributions from scalar perturbations in the shear, which give rise to source terms due to coupling
with density
perturbations. In this case $I_{a}$ satisfies a forced wave equation,
\begin{equation}
\label{eq:interaction}
\ddot{I}_{\langle a \rangle}- \dd^{2}I_{a} +3\Theta \dot{I}_{\langle a \rangle}
+\left[\frac{13}{9}\Theta^{2}
-\frac{1}{6}\mu-\frac{5}{2}\mu w+\frac{7}{3}\Lambda \right] I_{a} = \mathcal{C}^{I}_{a}\;,
\end{equation}
where the forcing term $\mathcal{C}^{I}_{a}$ is given by,
\begin{equation}
\mathcal{C}^{I}_{a}=-\frac{w}{a^{2}(1+w)}
\left(\dot{S}_{\langle
a\rangle} + \Theta S_{a} \right).
\end{equation}
To close the above system, we give the companion wave equation for $S_{a}=a\tilde{B}^{b}\dd_{\langle a}\mathcal{X}_{b
\rangle}$ as,
\begin{equation}
\label{eq:sawave}
\ddot{S_{a}} - w \dd^{2} S_{a}+(2-w)\Theta
\dot{S_{a}}+\left[\frac{2}{3}(1-w)(\Lambda+\Theta^{2})
+ \frac{1}{6}\mu(1+3w)(3w-5) \right]S_{a}=0.
\end{equation}
We note, for later convenience (\S~\ref{sec:bfield}) that the forcing term $\mathcal{C}^{I}_{a}=0$ in a matter dominated
universe ($w=0$) i.e $I_{a}$ decouples from $S_{a}$ when $w=0$.

\section{Induction of EM fields}

We introduce non-linear gravitationally induced `effective current' terms $\mathcal{C}^{E}_{a}$,
$\mathcal{C}^{\mathscr{E}}_{a}$ and
$\mathcal{C}^{\beta}_{a}$ which are made up of the coupling between density and gravitational wave
perturbations; these will act as driving forces of the induced Maxwell fields. 

\subsection{The Electric field}
\label{sec:efield}
We show how the coupling of gravitational perturbations with the seed magnetic field can induce Electric fields.
Here we give wave equations for the induced Electric field $E_{a}$ and its rotation $\mathscr{E}_{a}$.
In deriving the wave equation for $E_{a}$, we differentiate \ref{eq:induction_E} and equate the
result to the non-linear
identity,
\begin{equation}
\label{eq:identity}
\dot{(\mathrm{curl}\;B_{a})} = \mathrm{curl}\;\beta_{a}-\Theta\,
\mathrm{curl}\;B_{a}+H_{ab}\tilde{B}^{b}+\frac{1}{3a(1+w)}\epsilon_{abc}\tilde{B}^{b} \left(\Theta w
\mathcal{X}^{c}-2\dot{\mathcal{X}}^{c}\right)\;,
\end{equation}
obtained from the commutation relations (Appendix \ref{sec:crelations}) and we have used Equation \ref{eq:density} to rewrite the acceleration
terms.
The resulting wave equation is found to be,
\begin{eqnarray}
\label{eq:waveE}
\ddot{E}_{\langle a
\rangle}-\dd^{2}E_{a}+\frac{5}{3}\Theta\dot{E}_{a}+\left[\frac{2}{9}\Theta^{2}+\frac{1}{3}\mu(1-3w)+\frac{4}{3}\Lambda
\right] E_{a} &=&  \mathcal{C}^{E}_{a}\;,
\end{eqnarray}
where $\mathcal{C}^{E}_{a}$ is a gravitationally induced source term given by, 
\begin{equation}
\label{eq:sourceE}
\mathcal{C}^{E}_{a} = \mathrm{curl}\;I_{a}+ H_{ab}\tilde{B}^{b}
+\frac{1}{a(1+w)}\epsilon_{abc} \left[\left(w-\frac{2}{3} \right)(\tilde{B}^{b} \mathcal{X}^{c})\dot{} +
\Theta\left(w-\frac{4}{9} \right) \tilde{B}^{b} \mathcal{X}^{c}\right] -\Theta \mathcal{J}_{a} -\dot{\mathcal{J}_{a}}\;,
\end{equation}
and $\mathcal{J}_{a}$ is the 3-current. The terms involving $\epsilon_{abc}$ in $\mathcal{C}^{E}_{a}$ vanish when the
magnetic field
$\tilde{B}^{a}$ is parallel to the fractional
density gradient
$\mathcal{X}^{a}$. Taking the curl of \ref{eq:waveE} results in the equation governing the rotation of $E_{a}$,
\begin{eqnarray}
\label{eq:curlOfE}
\ddot{\mathscr{E}}_{a}-\dd^{2} \mathscr{E}_{a}+\frac{7}{3}\Theta
\dot{\mathscr{E}_{a}}+\left[\frac{7}{9}\Theta^{2} +\frac{1}{6}\mu(1-9w)+\frac{5}{3}\Lambda
\right]\mathscr{E}_{a} &=&  \mathcal{C}^{\mathscr{E}}_{a}\;,
\end{eqnarray}
where the source term $\mathcal{C}^{\mathscr{E}}_{a} = \mathrm{curl}\;\mathcal{C}^{E}_{a}$ is given by,
\begin{eqnarray}
\label{eq:sourceEc}
\mathcal{C}^{\mathscr{E}}_{a} &=& -(\mathrm{curl}\;\mathcal{J}_{a})\dot{}-\frac{4}{3}\Theta
\mathrm{curl}\;\mathcal{J}_{a} + 2\,\dd^{b}\dd_{[a}I_{b]}+\epsilon_{acd}\tilde{B}_{b}\dd^{c}H^{db} \nonumber \\
&+& \frac{2}{a^{2}(1+w)} \left[\left(w-\frac{2}{3}
\right) (a \tilde{B}_{[a} \dd^{b}\mathcal{X}_{b]})\dot{} + \Theta\left(w-\frac{4}{9} \right)
a \tilde{B}_{[a} \dd^{b} \mathcal{X}_{b]} \right]\;.
\end{eqnarray}

\subsection{The Magnetic field}
\label{sec:bfield}

As already mentioned, the induced magnetic field will be characterized via the variable
$\beta_{a}=\dot{B}_{a}+\sfrac{2}{3}\Theta B_{a}$.
On using \ref{eq:induction_B}, \ref{eq:interaction} and \ref{eq:curlOfE}, one can write a second-order equation
governing the evolution of the fundamental
variable $\beta_{a}$. This can be written in either of two forms: in terms of $I_{a}$ or $\mathscr{E}_{a}$,
corresponding to using \ref{eq:induction_B} as a constraint to either of \ref{eq:curlOfE} or \ref{eq:interaction} 
respectively.
Recall that both $I_{a}$ and $\mathscr{E}_{a}$ satisfy wave equations of the form
$\mathcal{L}[I_{a}]=\mathcal{C}^{I}_{a}$ and $\mathcal{L}[\mathscr{E}_{a}]=\mathcal{C}^{\mathscr{E}}_{a}$, where the
$\mathcal{C}^{i}_{a}$s are source terms.
 
Using covariant harmonics ~\cite{bruni:1992}, one can already notice from \ref{eq:interaction} and \ref{eq:curlOfE} that the eigenfunctions used to harmonically
 decompose
$I_{a}$ and $\mathscr{E}_{a}$ are not the same for a general perturbation \footnote{In
particular, for scalar
perturbations, we expand $I_{a}$ as
$I_{a}=\tilde{B}_{(n)}\sigma_{(k)}\tilde{\mathcal{H}}^{a}_{(n)}\dd_{\langle
a}\dd_{b \rangle}\mathcal{Q}^{(k)}$ and $\mathscr{E}_{a}$ as
$\mathscr{E}_{a}=\mathscr{E}_{(\ell)}\tilde{\mathcal{H}}^{(n)}_{[a}\dd^{b}\dd_{b]}\mathcal{Q}^{(k)}$; these are
evidently 
not the same eigenfunctions. Implicitly,
$\mathcal{C}^{I}_{a}$ will be expanded in the same harmonics as $I_{a}$, and similarly for $\mathscr{E}_{a}$ and
$\mathcal{C}^{\mathscr{E}}_{a}$.}. Consider the induction equation ~\ref{eq:induction_B}, and write it as $
\beta_{a}= \sum_{k}(\mathcal{P}_{a} I_{(k)} - \mathcal{Q}_{a} \mathscr{E}_{(k)})$, where $\mathcal{P}_{a}$ and
$\mathcal{Q}_{a}$
are distinct eigenfunctions of the Laplace-Beltrami operator, i.e $\mathcal{P}_{a} \neq \mathcal{Q}_{a}$. 
For the separation of variables technique to work for $\beta_{a}$, one must
eliminate either $I_{a}=\mathcal{P}_{a}
I_{(k)}$, 
along with its source terms $\mathcal{C}^{I}_{a}$ or $\mathscr{E}_{a}=\mathcal{Q}_{a} \mathscr{E}_{(k)}$ along with its
source terms $\mathcal{C}^{\mathscr{E}}_{a}$. In
this way, $\beta_{a}$ can then be expanded in terms of one set of complete eigenfunctions. This presents a problem:
since
 both $I_{a}$ and $\mathscr{E}_{a}$ are coupled to source terms $\mathcal{C}^{I}_{a}$ and
$\mathcal{C}^{\mathscr{E}}_{a}$ respectively at second-order, both $\mathcal{C}^{I}_{a}$ and
$\mathcal{C}^{\mathscr{E}}_{a}$
will still couple to the $\beta_{a}$ equation at this order, thereby introducing the differing set of eigenfunctions
$\mathcal{P}_{a}$ and $\mathcal{Q}_{a}$. A similar problem arose in \cite{ellis:1990}, due to the inclusion of a
vorticity term.

It is possible to do away with $\mathcal{C}^{I}_{a}$ in equation~\ref{eq:interaction} by requiring that
$w=0$ and this alleviates the
problem \footnote{This is not to say that $w=0$ is any more special than $w=1/3$, we simply invoke it here to decouple
~\ref{eq:interaction} from the source terms; indeed, any other method to achieve this would suffice. As a matter of
fact, when considering only tensor perturbations,~\ref{eq:interaction} does not couple to any source terms, even for a
general $w$. Moreover, when including vector perturbations, Equation~\ref{eq:interaction} does couple to a source term
$\sfrac{3}{2}\tilde{B}^{b}\dd_{\langle a}\dd^{c}\sigma_{b\rangle c}$, even when $w=0$.}. We shall then henceforth
restrict to the pressureless dust ($w=0$) case and write the $\beta_{a}$ wave equation in terms of
$\mathscr{E}_{a}$. 
\begin{eqnarray}
\label{eq:betawave}
\ddot{\beta}_{\langle a \rangle}-\dd^{2}\beta_{a}+3\Theta \dot{\beta}_{\langle a \rangle} +
\left[\frac{13}{9}\Theta^{2}-\frac{1}{6}\mu+\frac{7}{3}\Lambda \right]\beta_{a}&=& \mathcal{C}^{\beta}_{a} 
\end{eqnarray}
where, 
\begin{eqnarray}
\label{eq:sourcebeta}
\mathcal{C}^{\beta}_{a} &=& -\frac{2}{3}\Theta \dot{\mathscr{E}}_{a} +
 \left[-\frac{2}{3}\Theta^{2}+\frac{1}{3}\mu-\frac{2}{3}\Lambda
\right]\mathscr{E}_{a} + (\mathrm{curl}\;\mathcal{J}_{a})\dot{}+\frac{4}{3}\Theta\,
\mathrm{curl}\;\mathcal{J}_{a} - 2\,\dd^{b}\dd_{[a}I_{b]}-\epsilon_{acd}\tilde{B}_{b}\dd^{c}H^{db} \nonumber \\
&-& \frac{2}{a^{2}} \left[-\frac{2}{3}
 (a \tilde{B}_{[a} \dd^{b}\mathcal{X}_{b]})\dot{} -\frac{4}{9}\Theta
(a \tilde{B}_{[a} \dd^{b} \mathcal{X}_{b]}) \right]\;.
\end{eqnarray}
Note that while we keep $S_{a}=a\tilde{B}^{b}\dd_{\langle a}\mathcal{X}_{b \rangle}$ distinct from $a \tilde{B}_{[a}
\dd^{b} \mathcal{X}_{b]}$ in real space, their evolution equations can be made equivalent in harmonic space by a
suitable choice of eigenfunctions \footnote{In particular, expanding $a\tilde{B}^{b}\dd_{\langle a}\mathcal{X}_{b
\rangle}$ in terms of $a^{2}\tilde{\mathcal{H}}^{b}\dd_{\langle a}\dd_{b \rangle}\mathcal{Q}$ and $a \tilde{B}_{[a}
\dd^{b} \mathcal{X}_{b]}$ in terms of $a^2\tilde{\mathcal{H}}_{[a}\dd^{b}\dd_{b]}\mathcal{Q}$ will yield the
same harmonic components $S_{(\ell)}\equiv\tilde{B}_{(n)}\mathcal{X}_{(k)}$; $a^{2}\tilde{\mathcal{H}}^{b}\dd_{\langle
a}\dd_{b \rangle}\mathcal{Q}$ and $a^2\tilde{\mathcal{H}}_{[a}\dd^{b}\dd_{b]}\mathcal{Q}$ are eigenfunctions of the
Laplace-Beltrami operator.}. We shall thus write $S_{(\ell)}$ in place
of $\tilde{B}_{(n)}\mathcal{X}_{(k)}$ to avoid introducing another letter to denote the latter. This should not lead to
any ambiguities.
\subsection{The Electric Current}
\label{sec:current}

\subsubsection{Limiting cases: poor and perfect conductivity}
\label{subsub:limits}
To close the above system, one needs to take care of the current term $\mathcal{J}_{a}$ appearing in ~\ref{eq:sourceE}, ~\ref{eq:sourceEc} and
~\ref{eq:sourcebeta}. This term depends on the electrical properties of the medium. It is given in terms of the
Electric field $E_{a}$ via Ohm's law,
\begin{equation}
\label{eq:flOhm}
\mathcal{J}_{a}=\varsigma E_{a},
\end{equation}
where $\varsigma$ is the electrical conductivity of the medium. In this section, we consider only the
limiting cases of very high ($\varsigma\rightarrow \infty$) and very poor conductivity ($\varsigma\rightarrow 0$). 
Under the assumption of poor conductivity, the currents vanish $\mathcal{J}_{a}=0$, despite the presence of a non-zero
electric field. In
this case, one solves equations~\ref{eq:waveE}, \ref{eq:curlOfE} and \ref{eq:betawave}, with the current terms
 set to zero. At the opposite end, the case of perfect conductivity, the electric fields
vanish and the currents keep the magnetic field frozen in with the fluid. In this case, the
current term satisfies,
\begin{eqnarray}
\label{eq:current1}
(\mathrm{curl}\;\mathcal{J}_{a})\dot{}+\frac{4}{3}\Theta
\mathrm{curl}\;\mathcal{J}_{a} = 2\,\dd^{b}\dd_{[a}I_{b]}+\epsilon_{acd}\tilde{B}_{b}\dd^{c}H^{db} 
+ \frac{2}{a^{2}} \left[-\frac{2}{3}
 (a \tilde{B}_{[a} \dd^{b}\mathcal{X}_{b]})\dot{} -\frac{4}{9}\Theta
(a \tilde{B}_{[a} \dd^{b} \mathcal{X}_{b]}) \right],
\end{eqnarray}
and \ref{eq:waveE} and \ref{eq:curlOfE} are no longer relevant. One can verify that using this relation reduces
equation~\ref{eq:betawave} 
 to $\beta_{a}=I_{a}$, as can be confirmed also from the 
induction equation~\ref{eq:induction_B}.

One can also invoke the magnetohydrodynamic MHD approximation, which is valid for cold plasmas 
(pressureless dust can be well approximated by a cold plasma treatment) \cite{khanna:1998}  
. Cold
plasmas have components with non-relativistic velocities and are thus mathematically easier to deal with
\cite{Zunckel:2006cs,Kuroyanagi:2009ez,Marklund:2004qz}. We consider a two component electron-ion plasma and assume
that the motion properties of the plasma on macroscopic scales are captured by the center of mass 3-velocity $v^{a}$ of
 the system i.e the difference in mean velocities of the individual species is small compared with the fluid velocity.
We also assume charge neutrality of the cosmic plasma, i.e., the number densities of the electrons and ions $n_{e}$
and $n_{i}$ are roughly equal, $n_{e}\approx n_{i}$; this guarantees the vanishing of the total charge
$\rho_{c}=-e(n_{e}-n_{i})\approx 0$ and the background 3-current $\mathcal{J}_{a}\approx 0$.
In this case, the generalized Ohm's law is given by 
\begin{equation}
\label{eq:genOhm}
\mathcal{J}_{\langle a \rangle}=\varsigma (E_{a}+\epsilon_{abc}v^{b}\tilde{B}^{c})\,,  \qquad
v^{a}=\frac{\mu_{e} v_{e}^{a}+\mu_{i} v_{i}^{a}}{\mu_{e}+\mu_{i}}\;,
\end{equation}
 where the subscripts $e$ and $i$ denote quantities for electrons and ions respectively. The center of mass 3-velocity
$v_{a}$ of the electron-ion plasma can be
shown to satisfy the linearized Euler equation,
\begin{equation}
\label{eq:euler}
\dot{v}_{\langle a \rangle}+\frac{1}{3}\Theta v_{a}=0\;.
\end{equation}

In the ideal-MHD environment, the conductivity of the medium is very high ($\varsigma\rightarrow \infty$), then
$E_{a}+\epsilon_{abc}v^{b}\tilde{B}^{c}\rightarrow 0$ in order to keep the current $\mathcal{J}_{a}$ finite. This
readily gives the Electric field $E_{a}$ and its rotation $\mathscr{E}_{a}$ as,
$E_{a}=-\epsilon_{abc}v^{b}\tilde{B}^{c}$ and $
\mathscr{E}_{a} = 2 \tilde{B}_{[a} \dd^{b} v_{b]}$.
Using~\ref{eq:induction} and \ref{eq:euler}, one can show that,
\begin{equation}
\label{eq:MHDelec}
\dot{E}_{\langle a \rangle} + \Theta E_{a} =0, \qquad \text{and}\qquad
\dot{\mathscr{E}}_{a}+\frac{4}{3}\Theta\,\mathscr{E}_{a}=0\;.
\end{equation} 
With these, the 3-current $\mathcal{J}_{a}$ satisfies,
\begin{eqnarray}
\label{eq:current2}
(\mathrm{curl}\;\mathcal{J}_{a})\dot{}+\frac{4}{3}\Theta
\mathrm{curl}\;\mathcal{J}_{a} = 2\,\dd^{b}\dd_{[a}I_{b]}+\epsilon_{acd}\tilde{B}_{b}\dd^{c}H^{db} 
+ \frac{2}{a^{2}} \left[-\frac{2}{3}
 (a \tilde{B}_{[a} \dd^{b}\mathcal{X}_{b]})\dot{} -\frac{4}{9}\Theta
(a \tilde{B}_{[a} \dd^{b} \mathcal{X}_{b]}) \right]
 \nonumber \\
+\dd^{2}\mathscr{E}_{a}-\left(-\frac{1}{9}\Theta^{2}+\frac{5}{6}\mu+\frac{1}{3}\Lambda \right) \mathscr{E}_{a}\;.
\end{eqnarray}
Substituting \ref{eq:current2} into \ref{eq:betawave} results in
\begin{eqnarray}
\label{eq:betawave2}
\ddot{\beta}_{\langle a \rangle}-\dd^{2}\beta_{a}+3\Theta \dot{\beta}_{\langle a \rangle} +
\left[\frac{13}{9}\Theta^{2}-\frac{1}{6}\mu+\frac{7}{3}\Lambda \right]\beta_{a}&=&
\dd^{2}\mathscr{E}_{a}+\left[\frac{1}{3}\Theta^{2}-\frac{1}{2}\mu-\Lambda \right]\mathscr{E}_{a}\;.
\end{eqnarray}

The application of the ideal MHD approximation in cosmology has often been criticized as rather being of practical
appeal rather than of physical one \cite{Teodoro:2007pq}. Ideally, the curl of $E_{a}$ should be the outcome of a
rigorous treatment of the
physics of the particle interactions in terms of a kinetic theory description, see for example \cite{Ichiki10022006,Takahashi:2005nd}.

\subsubsection{Intermediate case: Finite conductivity}

The case of poor conductivity may not be much relevant in the post recombination epoch as the universe then acquires very high conductivity.
The perfect conductivity case, while relevant, may be thought of as an idealized notion. 
We thus turn to the finite conductivity case. The conductivity of the post decoupling era can be modelled by,
\begin{equation}
\varsigma = \frac{n_{e}^{2}e^{2}}{m_{e}n_{\gamma}\sigma_{T}}\approx 10^{11}\,\mathrm{s}^{-1}
\end{equation}
where, $n_{e}$ is the density of free electrons, $e$ is the electric charge of an electron, $m_{e}$ is the mass of an electron, $n_{\gamma}$ is the density of photons and $\sigma_{T}$ is the collision crossection. For a perfect fluid, the ratio $n_{\gamma}/n_{e}$ is constant, see \cite{battaner:2009} for example.

Assuming that Ohm's law holds (Equation \ref{eq:flOhm}), we may write the current terms in \ref{eq:sourcebeta} as,
\begin{eqnarray}
\label{eq:currentfinite}
(\mathrm{curl}\;\mathcal{J}_{a})\dot{}+\frac{4}{3}\Theta
\mathrm{curl}\;\mathcal{J}_{a} = \varsigma \dot{\mathscr{E}}_{a} + \frac{4}{3}\Theta \,\varsigma \mathscr{E}_{a}
\end{eqnarray}
where we have assumed that spatial gradients of the conductivity may be neglected $(\dd_{a}\varsigma \approx 0)$ and that the conductivity is constant 
in time ($\dot{\varsigma} \approx 0$). Substituting \ref{eq:currentfinite} in the wave equation~\ref{eq:betawave} for $\beta_{a}$ results in,
\begin{eqnarray}
\label{eq:betawavecond}
\ddot{\beta}_{\langle a \rangle}-\dd^{2}\beta_{a}+3\Theta \dot{\beta}_{\langle a \rangle} +
\left[\frac{13}{9}\Theta^{2}-\frac{1}{6}\mu+\frac{7}{3}\Lambda \right]\beta_{a}&=& \mathcal{C}^{\beta}_{a} 
\end{eqnarray}
where the source term $\mathcal{C}^{\beta}_{a}$ is now given by,
\begin{eqnarray}
\label{eq:sourcebetacond}
\mathcal{C}^{\beta}_{a} &=& \left( \frac{\varsigma}{\Theta}-\frac{2}{3}\right) \Theta \dot{\mathscr{E}}_{a} +
 \left[\left(2\frac{\varsigma}{\Theta} -1\right) \frac{2}{3}\Theta^{2}+\frac{1}{3}\mu-\frac{2}{3}\Lambda
\right]\mathscr{E}_{a} - 2\,\dd^{b}\dd_{[a}I_{b]}-\epsilon_{acd}\tilde{B}_{b}\dd^{c}H^{db} \nonumber \\
&-& \frac{2}{a^{2}} \left[-\frac{2}{3}
 (a \tilde{B}_{[a} \dd^{b}\mathcal{X}_{b]})\dot{} -\frac{4}{9}\Theta
(a \tilde{B}_{[a} \dd^{b} \mathcal{X}_{b]}) \right]\;.
\end{eqnarray}
%
Note that the electric currents $\mathcal{J}_{a}$, electric fields $E_{a}$ and the conductivity $\varsigma$ are all simultaneously finite. 
The simplifications that arise due to the characterisation of the limitting cases ($\mathcal{J}_{a}= 0$ for poor conducting mediums and  $E_{a}= 0$ for perfect conducting mediums) are no longer applicable in the case of finite conductivity. 
One then needs a proper model for the electric currents to ensure that the initial conditions for $\mathcal{J}_{a}$ and $E_{a}$ are not chosen independently. 
There are several ways in which one can model electric currents, all resulting in terms of perturbative order $\epsilon_{g}^{2}$, see \cite{Maeda:2008dv} for example. 
While these terms can be seamlessly accommodated in our framework, they have the undesirable effect of seeding magnetic fields.
This will lead us away from the isolated effects of the amplification of an already existing field. Inclusion of such terms will therefore lead us to overestimate the effect of the amplification. 
With this in mind, we restrict to the limitting cases of \ref{subsub:limits}.

\section{The Induced Fields}

We now treat separately the induction of electromagnetic fields due to
interaction with \textit{scalar} and \textit{tensor} perturbations. To this end, we expand the perturbation 
variables in terms of a helicity basis (Appendix~\ref{sec:harmonics}). In addition, 
we use a unified time variable whose defining equation is
$\dot{\tau}=\sfrac{3}{2}H_{i}$ instead of
proper time, to re-write the relevant equations \footnote{The subscript $i$ marks some initial time $\tau=1$, see
\cite{Betschart:2005iu} for details.}.
We have to substitute
for $\mu$, $\Theta$ and $a$, appearing in the perturbed equations, from the zeroth order equations. We restrict our
treatment to zero cosmological constant $\Lambda=0$ and flat spatial sections $K=0$. Friedmann equation
then reduces to $\mu=\Theta^{2}/3$, where $\Theta$ is given by $\Theta=3H_{i}/\tau$; the scale factor $a$
evolves as
$a=a_{i}\tau^{2/3}$.

\subsection{EM induction due to scalar perturbations}

In this case, the coupling of a seed field with gravitational perturbations is described by the variable $I_{a}$ and
$S_{a}$; these variables become sources of electromagnetic fields. 

\begin{itemize}
\item{\textbf{Interaction terms}}: Equations~\ref{eq:interaction} and \ref{eq:sawave} for the interaction variables $I_{a}$ and $S_{a}$, respectively become,
\begin{subequations}
\begin{equation}
\label{har:interaction}
\frac{9}{4}I''_{(\ell)}+\frac{27}{2\tau}I'_{(\ell)}+\frac{25}{2\tau^{2}} I_{(\ell)} = 0, \\
\end{equation}
\begin{equation}
\frac{9}{4}S''_{(\ell)}+\frac{9}{\tau}S'_{(\ell)}+\frac{7}{2\tau^{2}} S_{(\ell)}=0\;,
\end{equation}
\end{subequations}
Note that since $w=0$, the entire system has decoupled from $a\tilde{B}^{b}\dd_{\langle a}\mathcal{X}_{b \rangle}$, however we
still need an equation for $S_{(\ell)}$ because of the coupling with $a \tilde{B}_{[a}
\dd^{b} \mathcal{X}_{b]}$ in Equations \ref{eq:sourceEc} and \ref{eq:sourcebeta}. 
These interaction variables have the general solutions,
\begin{equation}
I_{(\ell)}(\tau)=C_{1}\tau^{-10/3}+C_{2}\tau^{-5/3} \qquad \text{and} \qquad
S_{(\ell)}=\frac{1}{5}C_{3}\tau^{-7/3} +\frac{1}{5}C_{4}\tau^{-2/3}\;,
\end{equation}
where the $C_{i}$'s are integration constants.

\item{\textbf{EM fields}}: 
%
%
Equation~\ref{eq:waveE} for the Electric field $E_{a}$ becomes,
%
\begin{equation}
\label{har:electric}
\frac{9}{4}E''_{(\ell)}+\frac{15}{2\tau}E'_{(\ell)}+\left[\left(\frac{\ell}{a_{i}
H_{i}}\right)^{2}\tau^{-4/3} +\frac{3}{\tau^{2}}  \right] E_{(\ell)}=\pm \frac{(k+n)}{3a_{i}H^{2}_{i}}
I_{(\ell)}\tau^{-2/3} \mp \frac{1}{H_{i}}S_{(\ell)}'
\mp \frac{4}{3H_{i}\tau}S_{(\ell)}
\end{equation}
%
%
%
It is much easier to solve for $\beta_{(\ell)}$ from the induction equation 
\begin{equation}
\beta_{(\ell)} = I_{(\ell)}\mp \frac{\ell}{a_{i}\tau^{2/3}} E_{(\ell)} 
\end{equation}
 once $I_{(\ell)}$
 and $E_{(\ell)}$ are known, rather than from the wave equation~\ref{eq:betawave}.

%
%
%
\end{itemize}

\subsection{EM induction due to tensor perturbations}

In this case, the transverse and trace-free parts of the shear tensor $\sigma_{ab}$ characterize gravitational waves.
The interaction with a
seed field is then purely described by the variable $I_{a}$ without any contribution from either density or velocity
perturbations. The generalized Ohm's law~\ref{eq:genOhm} in the MHD approximation also reduces to ~\ref{eq:flOhm}.
We thus only need the equations for $\beta_{a}$, $I_{a}$ and $E_{a}$.
\begin{itemize}
\item{\textbf{Interaction variable}}: Equation~\ref{eq:interaction} for the interaction variable $I_{a}$ becomes,
\begin{eqnarray}
\label{har:interactionT}
\frac{9}{4}I''_{(\ell)}+\frac{27}{2\tau}I'_{(\ell)}+\left[\left(\frac{\ell}{a_{i}
H_{i}}\right)^{2}\tau^{-4/3} +\frac{25}{2\tau^{2}} \right]I_{(\ell)} =0\;,
\end{eqnarray}
with the general solution,
\begin{equation}
I_{(\ell)}(\tau) =  \tau^{-5/2} \left[C_{1}J_{1}\left({\frac{5}{2}},\frac{\ell}{a_{i}H_{i}} \frac{2}{\tau^{1/3}}\right)
+ C_{2}J_{2}\left({\frac{5}{2}},\frac{\ell}{a_{i}H_{i}} \frac{2}{\tau^{1/3}}\right) \right]\;,
\end{equation}
where $C_{1}$ and $C_{2}$ are integration constants, $J_{1}$ and $J_{2}$ are Bessel functions of the second kind.

\item{\textbf{EM Fields}}: Equation ~\ref{eq:waveE} for the electric field variable $E_{a}$
becomes,
%
%
\begin{equation}
\label{har:electricT}
\frac{9}{4}E''_{(\ell)}+\frac{15}{2\tau}E'_{(\ell)}+\left[\left(\frac{\ell}{a_{i}
H_{i}}\right)^{2}\tau^{-4/3} +\frac{3}{\tau^{2}}  \right] E_{(\ell)} =
\pm \frac{(2k+n)}{H^{2}_{i}a_{i}}I_{(\ell)}\tau^{-2/3}
\end{equation}
and we once again determine $\beta_{(\ell)}$ from 
\begin{equation}
\beta_{(\ell)} = I_{(\ell)}\mp \frac{\ell}{a_{i}\tau^{2/3}} E_{(\ell)} 
\end{equation}
instead of using the wave equation \ref{eq:betawave}.
\end{itemize}

\section{Initial Conditions}

We need initial conditions in order to solve the equations in the previous section. The conditions are adapted 
as follows: for $\beta_{a}$ we invoke Maxwell's equation~\ref{eq:induction_B}
\begin{eqnarray}
\beta_{a} = I_{a} - \mathscr{E}_{a}, && \qquad \dot{\beta}_{a} = \dot{I}_{a} - \dot{\mathscr{E}}_{a} 
\end{eqnarray}
For the interaction variable $I_{a}$, we use the definition \ref{eq:defs} and Equation~\ref{eq:induction}
\begin{equation}
I_{a} = \sigma_{ab}\tilde{B}^{b}  \qquad \dot{I}_{a} = \dot{\sigma}_{ab}\tilde{B}^{b} + \sigma_{ab}\dot{\tilde{B}}^{b} \qquad \dot{\tilde{B}}_{a} = -\frac{2}{3}\Theta \tilde{B}_{a}
\end{equation}
For the rotation of the Electric field $\mathscr{E}_{a}$ \footnote{Note that the Electric field $E_{a}$ and its 
curl $\mathscr{E}_{a}$ are related by a factor of $\ell$ when expanded in terms of the helicity basis described in Appendix~\ref{sec:harmonics}. 
In particular $\mathscr{E}_{a}=\pm (\ell/a_{i}) \tau^{-2/3} E^{\scriptstyle{(\pm)}}e^{\scriptstyle{(\pm)}}_{a}$}, we use Maxwell's equation \ref{eq:induction_E} and the commutation relation~\ref{eq:ccurl} to get,
\begin{eqnarray}
\dot{\mathscr{E}}_{a} &=& -\Theta \mathscr{E}_{a}+\mathcal{R}_{ab}\tilde{B}^{b}-\dd^{2}B_{a} 
\end{eqnarray}
where in this case $B_{a}$ (without the tilde) is the induced magnetic field, 
and we have written the first order perturbed 3-ricci tensor $\mathcal{R}_{ab}$ as \cite{ellis:1971,Ellis:1998ct}
\begin{equation}
\mathcal{R}_{ab} = -\dot{\sigma}_{\langle ab \rangle} - \Theta \sigma_{ab}
\end{equation}
We require that the gravitationally induced field variables $E_{a}$ (and hence $\mathscr{E}_{a}$) and $B_{a}$ be zero initially.  This
leads to the following initial conditions for the perturbation variables:
\begin{equation}
\label{eq:inits}
\begin{array}{cclrcl}
I^{i}_{(\ell)} &=& \sigma^{i}_{(k)} \tilde{B}^{i}_{(n)} &\qquad \qquad I'_{i(\ell)} &=& \sigma'_{i(\ell)}\tilde{B}^{i}_{(n)}-\frac{4}{3} \sigma^{i}_{(k)}\tilde{B}^{i}_{(n)} \\
\mathscr{E}^{i}_{(\ell)} &=&  0 &       \mathscr{E}'_{i(\ell)} &=& -2 \mathscr{E}^{i}_{(\ell)} -\left(\sigma'_{i(\ell)}+
2\sigma^{i}_{(k)}\right) \tilde{B}^{i}_{(n)} \\
\beta^{i}_{(\ell)} &=& \sigma^{i}_{(k)} \tilde{B}^{i}_{(n)} &   \beta'_{i(\ell)} &=&
2\sigma'_{i(k)}\tilde{B}^{i}_{(n)}+\frac{2}{3}\sigma^{i}_{(k)} \tilde{B}^{i}_{(n)}+2\mathscr{E}^{i}_{(\ell)}
\end{array}
\end{equation}
%
%
Following \cite{Tsagas:2001ak,Tsagas:2005ki,Kuroyanagi:2009ez}, we adopt the initial
condition for the shear from $(\sigma / H)_{i}\sim 10^{-6}$. We choose the seed field to be $\tilde{B}^{i}=10^{-20} \;\mathrm{G}$,
as typical of those produced around the recombination era \cite{Takahashi:2005nd}. 

\section{Results}
\label{sec:powerspec}

\begin{figure}[h!]
\centering
\includegraphics[width=0.7\textwidth,height=8.5cm]{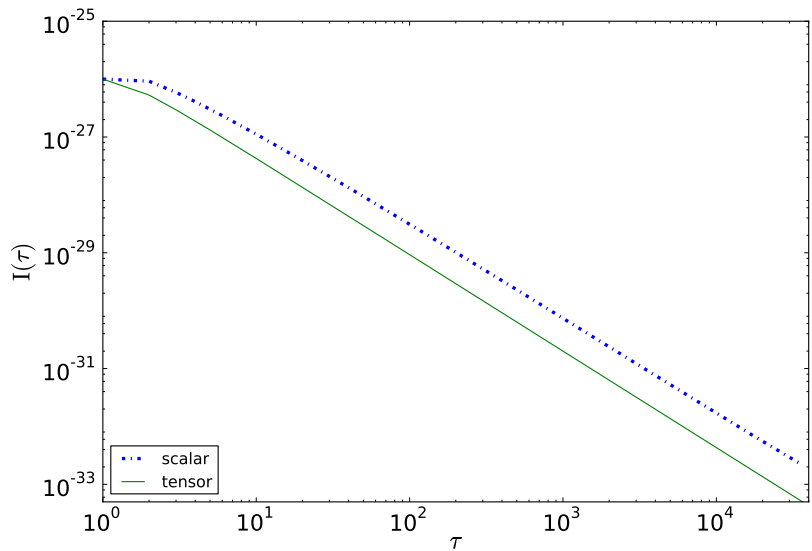}
\caption{\small{\textit{Time evolution of the interaction variable in log-log axes. 
Note that for the interaction with scalars, the decay is slightly slower than for tensors.}}}
\label{fig:ia}
\end{figure}

Given the system of initial conditions \ref{eq:inits}, one can notice that the interaction variable $I_{a}$ plays the 
fundamental role in the interaction process. If we set $I_{a}=0$ initially, then no amplification takes place.
We show the time evolution $I_{a}(\tau)$ in Figure~\ref{fig:ia} on a log-log scale. A noteworthy feature is the rapid decay of $I_{a}$ for both scalars and tensors. Although the interaction with scalar perturbations decays slightly slower, it essentially 
follows the same trend as the interaction with gravitational waves. We are thus led to conclude that even including scalar perturbations in the interaction,
 we reach the same conclusion as \cite{Fenu:2008vs} and \cite{Kuroyanagi:2009ez} that there is no significant amplification of electromagnetic fields coming from the interaction.

\begin{figure}[h!]
\centering
 \subfigure[]{\label{fig:tensor}\includegraphics[width=0.49\textwidth,
  height=7cm]{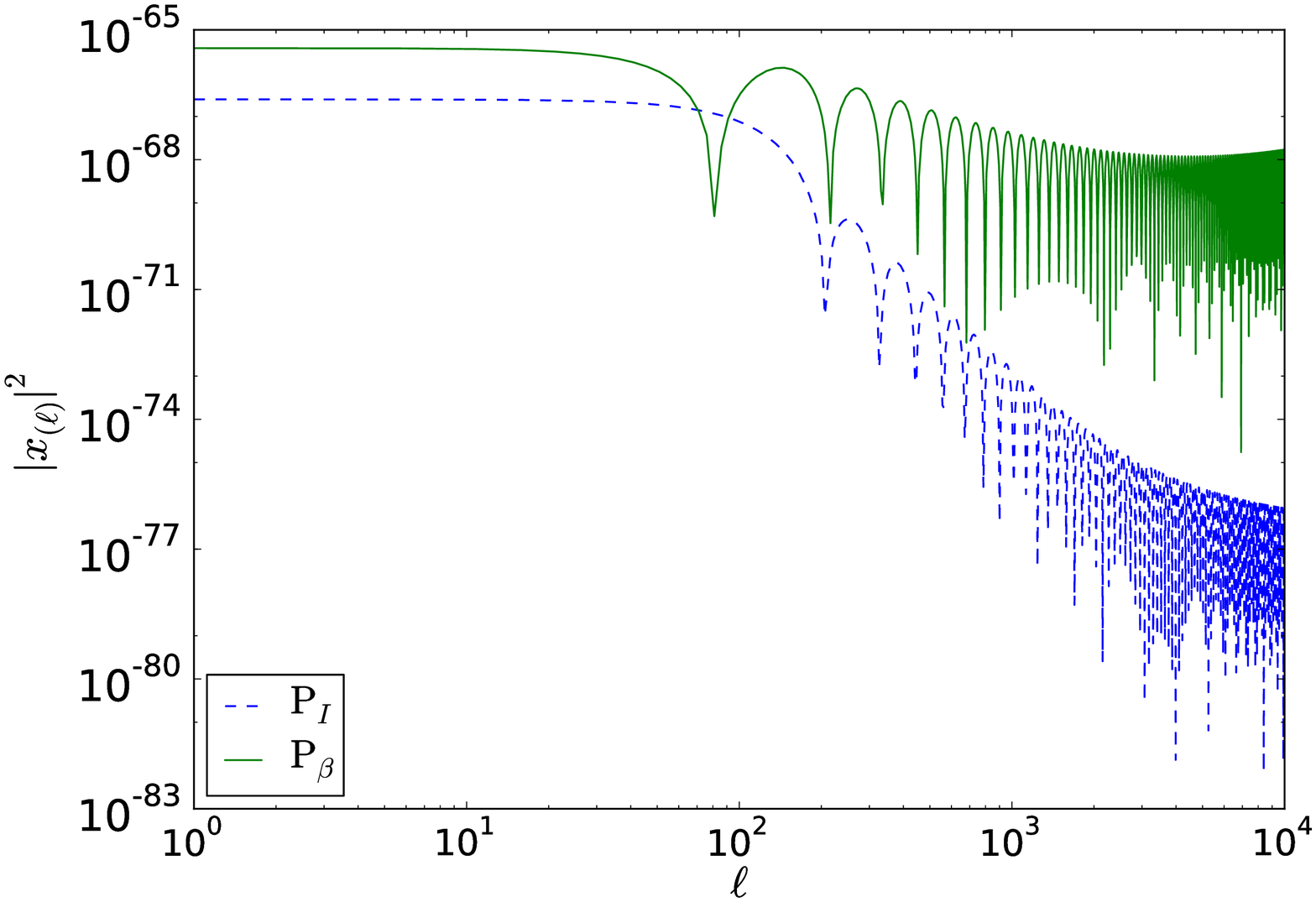}}
  \subfigure[]{\label{fig:scalar}\includegraphics[width=0.49\textwidth,
  height=7cm]{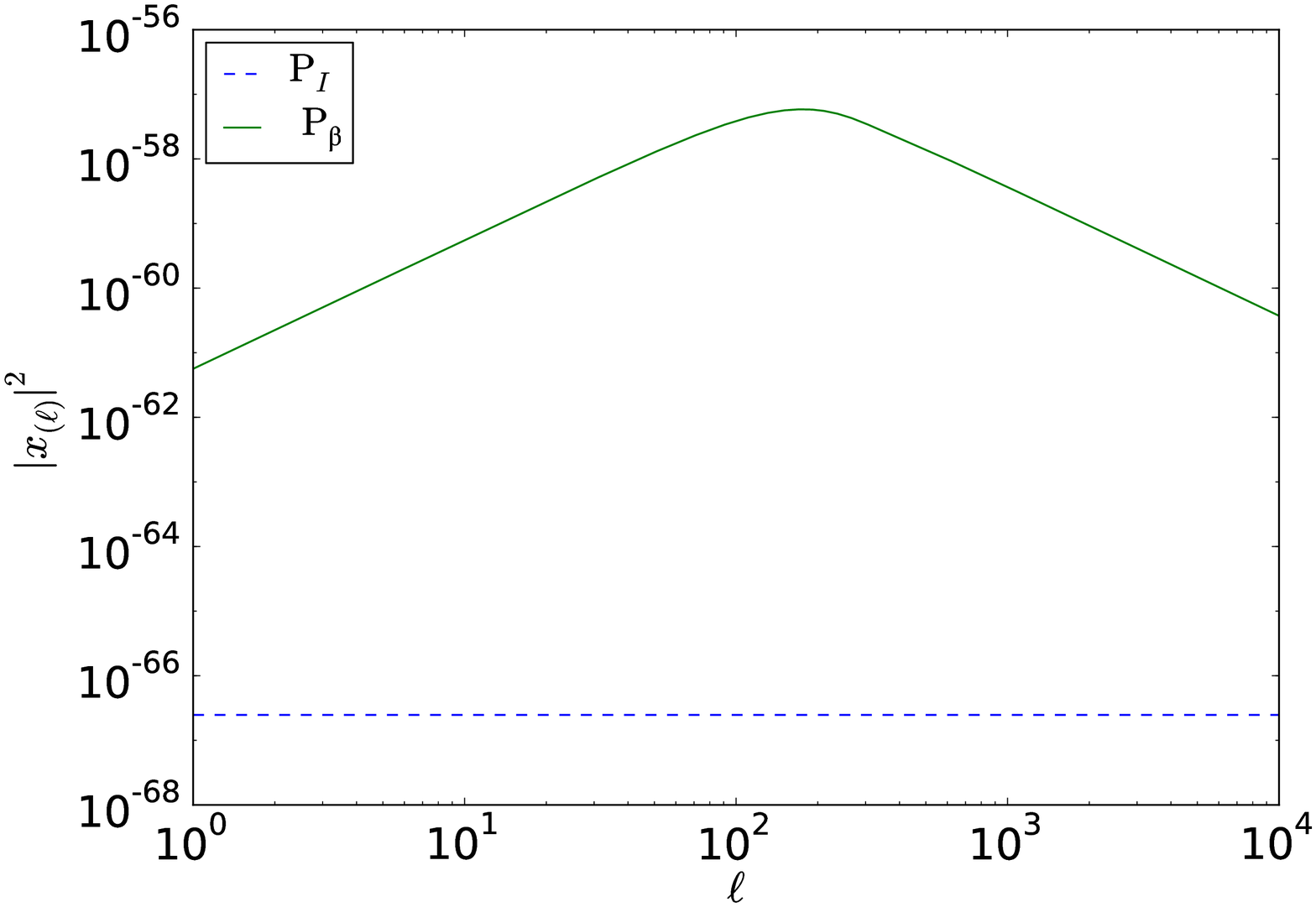}}
\caption{\small{\textit{Plots of Power vs scale ($\ell$); we define the power
as 
$P_{x}=|x(\ell)^{2}|$. (a): Power spectra of the magnetic field variable $\beta_{(\ell)}$ (green, solid),
 and the interaction variable $I_{(\ell)}$ (blue, dashed) at redshift $z=0$ for the
tensor case. (b): Power spectra of the magnetic field variable $\beta_{(\ell)}$ (green, solid),
 and the interaction variable $I_{(\ell)}$ (blue, dashed) at redshift $z=0$ for the
scalar case. }}}
\end{figure}

The effect of the gravitational perturbations on the interaction is thought to be largest at the point where the
modes enter the
horizon. This is clearly evident in Figures \ref{fig:tensor} and \ref{fig:scalar}.
A couple of features are worth noting from Figure~\ref{fig:tensor}. One is that the spectrum for the interaction
variable mimics that of gravitational waves. It is also consistent with the fact that gravitational waves 
start oscillating at horizon crossing. This is to be expected since although for a spatially  inhomogeneous 
magnetic field $\tilde{B}_{a}$, the product $I_{(\ell)} = \tilde{B}_{(n)} \sigma_{(k)}$ becomes a convolution in 
Fourier space, $I(k)=\sum_{n}B(n)\sigma(k-n)$, we have only considered the mode-mode coupling case, $I(k)=B(k)\sigma(k)$.

The power spectra for the case of interaction with scalars are not as interesting. There is no
scale dependence on the interaction variable $I_{a}$, cf equation \ref{har:interaction}. This is because the Laplacian
term for scalar perturbations comes from the acceleration vector which is identically zero in the dust case
$\mathcal{A}=0$. 

It would be interesting to generalize our treatment to include the case of non-zero pressure. This will lend us
to the radiation dominated era where one can incorporate photons in the plasma and can consider collisional effects as
was done in \cite{Ichiki10022006,Takahashi:2005nd} for example. One could treat the interesting case of simultaneous generation
and amplification of magnetic fields by coalescing these phenomenon.

\section{Conclusions}
\label{sec:conclusion}

We have carried out an analysis of the coupling between gravitational perturbations with electromagnetic fields
 as a possible means for magnetic field amplification. This carries to completion the work began in
\cite{Betschart:2005iu} and \cite{Zunckel:2006cs}.
In agreement with the work of \cite{Fenu:2008vs} and \cite{Kuroyanagi:2009ez} we argue that there is no significant
amplification resulting from the interaction of magnetic field with gravitational waves. Even with the inclusion of
density perturbations, the induced fields may still be orders of magnitude smaller. This justifies the perturbative
treatment and our neglect of
backreaction.

The induction of electromagnetic fields due to the interaction of a test magnetic field with gravitational waves was
 studied in \cite{Marklund:2000zs} using the weak-field approximation. We included this
study here treating the
background magnetic field as a first order perturbation and recovered similar results. This shows that there is no
fundamental difference between the two approaches, apart from a labeling of spacetimes, which should not affect
physical results. We also extended this study by using a proper non-linear perturbative framework. This framework was
applied in \cite{Betschart:2005iu}, but an erroneous argument there led to the neglect of the rotation
of the Electric field, thus restricting the study to perfectly conducting environments. This was refuted in \cite{Tsagas:2005qx}. 
In fact, upon inspection of \ref{eq:inits} one can conclude that even if one initially sets the rotation of the Electric field to
zero, $\mathscr{E}_{i}=0$ there are non-zero terms on the right hand side of the initial conditions for $\dot{\mathscr{E}}$ that
will seed a non-zero $\mathscr{E}$.
We also carry to completion the work in \cite{Zunckel:2006cs} by doing a proper extraction of the scalar and tensor
modes and numerical integrations. In terms of the conductivity of the cosmic medium, \cite{Marklund:2000zs} restricted
their study to poor conducting mediums, \cite{Betschart:2005iu} to perfectly conducting mediums and \cite{Zunckel:2006cs}
treated the MHD approximation. We carried our analysis for all three cases. We find that for tensor perturbations, the
ideal MHD approximation is just the same as the perfect conductivity assumption of the fluid treatment. For scalar
perturbations, we find an
additional source term in the induced field (compared with perfectly conducting environments) due to the coupling of the
seed field with scalar velocity perturbations.
The current term $\mathcal{J}_{a}$ was neglected at all orders in \cite{Betschart:2005iu}, in an attempt
presumably to uphold the background magnetic field's homogeneity condition $\dd_{a}\tilde{B}_{b}=0$. However, this is
not necessary since introducing the current term at the
non-linear order does not break the condition $\dd_{a}\tilde{B}_{b}=0$. Also, one cannot consistently invoke Ohm's law
for poor and perfect conducting environments without a current term. In \cite{Zunckel:2006cs}, an inhomogeneous seed
field was assumed thereby requiring a first order current $\mathcal{J}^{a}=\rho_{e}v_{e}^{a}+\rho_{i}v_{i}^{a}=-e(n_{e}
v_{e}^{a} -n_{i}v_{i}^{a})$ to
uphold the condition $\dd_{a}\tilde{B}_{b}\neq 0$. However, after decoupling (which is the era
considered there), Thompson scattering is no longer efficient. Thus electrons and ions are
tightly coupled by coulomb scattering at first order. Their velocity fields are therefore equal as they form a
perfectly coupled baryon fluid \cite{MNR:MNR9442,Matarrese:2004kq}. There can be therefore no currents generated at this
order and
the condition
$\mathrm{curl}\;\tilde{B}_{a}=\mathcal{J}_{a}$ will render the seed field homogeneous.

Both \cite{Betschart:2005iu} and \cite{Zunckel:2006cs} integrate $\beta_{a}$ to recover the amplified magnetic
field, after specifying a frame $u^{a}$. While this takes into account the frame dependence of the magnetic field
$B_{a}$, it invalidates gauge invariance as the recovered $B_{a}$ remains gauge dependent and takes the same value and
form as it would have without the introduction of $\beta_{a}$. This is already pointed out in \cite{Tsagas:2005qx},
See also \cite{Betschart:2007ts} and \cite{Tsagas:2007ux}.  We do not
integrate $\beta_{a}$ but simply note that one can assign a physical meaning to the
magnetic field variable $\beta_{a}$ by noting that $\beta_{a}=0$ describes the background adiabatic decay of the fields.
Any deviation from $\beta_{a}=0$ would then imply amplification of the background field. Moreover, $\beta_{a}$ is a
linear combination of terms that source magnetic fields through the induction equation \ref{eq:induction_B}. Thus we can
estimate the relative importance of each source term through $\beta_{a}$ without having to integrate it to recover the
gauge-dependent $B^{a}$. For example, we see from Figure~\ref{fig:tensor} that the rotation of the electric field
dominates at
small scales compared to the interaction term. Observations of cosmological magnetic fields are difficult enough as it
is, a new cosmological observable would lead to better understanding of studies in magnetic fields. While $\beta_{a}$
may not be that quantity, it does arise naturally from Maxwell's equations.


Also, one can readily write our key equations in terms of metric variables by adoption of a suitable tetrad as was done in \cite{Fenu:2010kh}.

Mechanisms that seek to generate magnetic fields, relying on non-linear perturbation theory are attractive for several
reasons \cite{Durrer10022006}. Among these
is that they can easily blend in with known physics as they become relevant around the recombination era. This
makes it possible to quantitatively evaluate the generated fields using CMB constraints. Progress in non-linear
perturbation theory will allow us to investigate these non-linear effects in a manner that is free of
spurious gauge modes \cite{Clarkson:2003af,Clarkson:2011qk}.

\section{Acknowledgments}
BM acknowledges financial support from the National Research Foundation. We thank Chris Clarkson, Roy Maartens and
Obinna Umeh for valuable comments. The analytic computation in this paper was performed with the help of the computer
package xAct \cite{xperm:2008}.

\appendix
\section{Appendix}
\subsection{Harmonic Splitting}

\label{sec:harmonics}
It is standard to decompose the perturbed variables harmonically in Fourier space; separating out the time 
and space variations \cite{harrison:1967,abbott:1986,bruni:1992}. The
idea is to expand the quantities in terms of eigenfunctions of the Laplace-Beltrami operator. To this end, 
we introduce the helicity basis vectors $\mathbf{e}^{\scriptscriptstyle{(-)}}, \mathbf{e}^{\scriptscriptstyle{(0)}}$ and  
$\mathbf{e}^{\scriptscriptstyle{(+)}}$ defined by

\begin{equation}
e^{\scriptscriptstyle{(\pm)}}_{a} = - \frac{i}{\sqrt{2}}\left(e^{1}_{a}\pm i e^{2}_{a}\right)
\end{equation}
where $(\mathbf{e^{1}}, \mathbf{e^{2}}, \mathbf{\hat{k}})$ form a right-handed orthornomal system with 
$\mathbf{e_{2}=\mathbf{\hat{k}} \times \mathbf{e}_{1}}$ and we align $\mathbf{e}^{\scriptscriptstyle{0}}$ with $ \mathbf{\hat{k}}$.

Using this basis, the scalar harmonic functions are given by,

\begin{eqnarray}
Q^{(0)} = e^{ik_{j}x^{j}} \;.
\end{eqnarray}
Scalar type components of vectors and tensors are expanded in terms of harmonic functions defined from $Q^{(0)}$ as follows,
\begin{eqnarray}
Q^{(0)}_{a} &=& -\frac{a}{k}\dd_{a}Q^{(0)}=a\,i\hat{k}_{a}e^{ik_{j}x^{j}} \, ,\\
Q^{(0)}_{ab} &=& \frac{a^{2}}{k^{2}}\dd_{\langle a}\dd_{b \rangle} Q^{(0)}  = -a^{2}\,\left(\hat{k}_{a}\hat{k}_{b}-\frac{1}{3}\delta_{ab} \right)e^{ik_{j}x^{j}}\,.
\end{eqnarray}
Vector harmonics are given by
\begin{eqnarray}
Q^{\scriptscriptstyle{(\pm)}}_{a} &=& e^{\scriptscriptstyle{(\pm)}}_{a}Q^{(0)}\; ,\\
Q^{\scriptscriptstyle{(\pm)}}_{ab} &=& -\frac{a}{k} \dd_{(a}e^{\scriptscriptstyle{(\pm)}}_{b)}Q^{(0)} = a\,i\hat{k}_{(a} e^{\scriptscriptstyle{(\pm)}}_{b)}e^{ik_{j}x^{j}}\;.
\end{eqnarray}
While tensor harmonics are defined as,
\begin{equation}
Q^{\scriptscriptstyle{\pm 2}}_{ab} = \sqrt{\frac{3}{2}} e^{\scriptscriptstyle{(\pm)}}_{a}e^{\scriptscriptstyle{(\pm)}}_{b}Q^{(0)}\;.
\end{equation}

%


\subsection{Maxwell's Equations}
\label{sec:maxwell}
The Maxwell field tensor $F_{ab}$ decomposes relative to the fundamental observer as,
\begin{equation}
F_{ab}=2u_{[a}E_{b]}+\epsilon_{abc}B^{c}\;,
\end{equation}
where  $E_{a}=F_{ab}u^{b}$ and $B_{a}=\frac{1}{2}\epsilon_{abc}F^{bc}$ are respectively the Electric and Magnetic field
as measured
by the fundamental observer moving with 4-velocity $u^{a}$. These are 3-vectors on the spacelike hypersurface,
$E_{a}u^{a}=0=B_{a}u^{a}$.
The Maxwell's equations are given by
\begin{equation}
\nabla_{[a}F_{bc]}=0\quad \mathrm{and}\quad \nabla^{b}F_{ab}=J_{a}\;,
\end{equation}
where $J$ is the 4-current. These equations can be decomposed
covariantly into the
following \cite{Tsagas:2004kv,Barrow:2006ch,ellis:1971}
\begin{subequations}
\label{eq:max}
\begin{equation}
\label{eq:max1}
\dot{E}_{\langle a \rangle} - \mathrm{curl}\; B_{a} = -\frac{2}{3}\Theta E_{a} + \sigma_{ab} E^{b} + \epsilon_{abc}
(\mathcal{A}^{b}
B^{c}+\omega^{b} E^{c}) -\mu_{0} \mathcal{J}_{\langle a \rangle}\;,
\end{equation}
\begin{equation}
\label{eq:max2}
\dot{B}_{\langle a \rangle} + \mathrm{curl}\; E_{a} = -\frac{2}{3}\Theta B_{a} + \sigma_{ab} B^{b} + \epsilon_{abc}
(\mathcal{A}^{b}
E^{c}+\omega^{b} B^{c})\;,
\end{equation}
\begin{equation}
\label{eq:max3}
0 = D_{a} E^{a} - 2\omega_{a} B^{a}- \frac{\rho_{c}}{\epsilon_{0}}\;,
\end{equation}
\begin{equation}
\label{eq:max4}
0 = D_{a} B^{a} + 2 \omega_{a} E^{a}\;.
\end{equation}
\end{subequations}
The EM fields are solenoidal in the absence of gravitational vector perturbations.  

\subsection{Commutation Relations}
\label{sec:crelations}
\begin{eqnarray}
(\dd_{a}f)\dot{}_{\perp} &=& \dd_{a}\dot{f}-\frac{1}{3}\Theta \dd_{a}f+\dot{f}\mathcal{A}_{a}\;, \\
(\dd_{a} V_{b})\dot{}_{\perp} &=& \dd_{a} \dot{V}_{b}-\frac{1}{3}\Theta \dd_{a}
V_{b}-\sigma_{a}^{\phantom{c}c}\dd_{c}V_{b}+\epsilon_{bcd}H_{a}^{\phantom{c}c}V^{d} +\mathcal{A}_{a}\dot{V}_{b}\;,\\
\label{eq:ccurl}
(\mathrm{curl}\,V_{a})\dot{}_{\perp} &=&
\mathrm{curl}\,\dot{V}_{a}-\frac{1}{3}\Theta\,\mathrm{curl}\,V_{a}-\epsilon_{abc}
\sigma^{bd} \dd_{d} V^{c} +H_{ab}V^{b} -\frac{1}{3}\Theta \epsilon_{abc}V^{b}\mathcal{A}^{c}\;, \\
\mathrm{curl}\;\mathrm{curl}\;S_{ab}&=&
-\dd^{2}S_{ab}+(\mu+\Lambda-\frac{1}{3}\Theta^{2})S_{ab}+\frac{3}{2}\dd_{\langle a}\dd^{c}S_{b\rangle c}\;.
\end{eqnarray}


\end{document}